\documentclass[a4paper,amsmath,amssymb]{revtex4} 

\usepackage{amsmath}
\usepackage{graphicx}

\begin{document}

\title{Attraction controls the entropy of fluctuations in isosceles
  triangular networks}

\author{Fabio Leoni}
\author{Yair Shokef\footnote{\texttt{shokef@tau.ac.il}}}
\affiliation{School of Mechanical Engineering and Sackler Center for
  Computational Molecular and Materials Science, Tel-Aviv University, 
  Tel-Aviv 69978, Israel} 

\begin{abstract}
We study two-dimensional triangular-network models, which have
degenerate ground states composed of straight or randomly-zigzagging
stripes and thus sub-extensive residual entropy. 
We show that attraction is responsible for the inversion of the stable
phase by changing the entropy of fluctuations around the
ground-state configurations. 
By using a real-space shell-expansion method, we compute the exact
expression of the entropy for harmonic interactions, while for
repulsive harmonic interactions we obtain the entropy arising from a
limited subset of the system by numerical integration.  
We compare these results with a three-dimensional triangular-network
model, which shows the same attraction-mediated selection mechanism of
the stable phase, and conclude that this effect is general with respect
to the dimensionality of the system.
\end{abstract}

\maketitle

\section{Introduction}

Geometrically-constrained systems may show peculiar features compared 
to their unconstrained counterparts. In particular, geometric
constraints can lead to frustration because the system cannot
simultaneously minimize all local interaction energies, or free
energies. 
Frustrated systems usually show a degenerate ground state and then
they may posses a residual entropy.  
Frustration is relevant in physical and biological systems that range
from water \cite{pauling1935} and spin ice \cite{harris1997,libal2018}
to magnets \cite{bramwell2001}, magnetic island \cite{wang2006}, high 
transition-temperature superconductors \cite{anderson1987}, elastic
beams \cite{kang2014,coulais2016} and colloids 
\cite{han2008,libal2006,ortiz2016,tierno2016}.  
The possibility to control colloidal interparticle interactions and to
visualize and manipulate each particle and follow its motion in both
space and time makes colloidal suspension a powerful tool to study
phenomena in condensed-matter physics, ranging from glass formers
\cite{gokhale2016}, to crystals and gels \cite{lu2013}.

\begin{figure}[h!]
\centering
\includegraphics[clip=true,width=17cm]{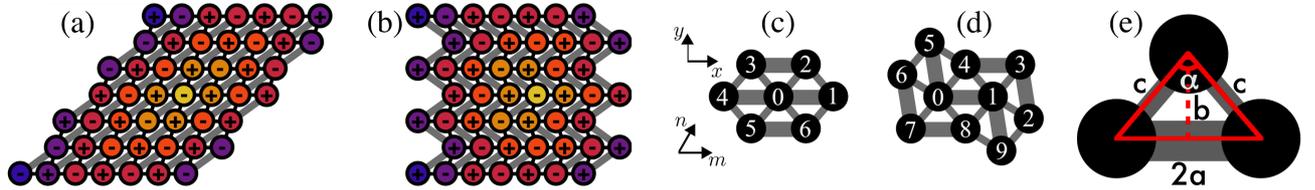}
\caption{(\textbf{a}) Straight and (\textbf{b}) maximally zigzagging
  or bent configuration. Shells, which order is indicated with $n_s$,
  are denoted by increasingly darker colors for increasing
  $n_s$. Thicker, gray lines correspond to springs of longer (for
  $\alpha>\pi/3$) or shorter (for $\alpha<\pi/3$) length at
  rest. (\textbf{c}) and (\textbf{d}) are the unit cell for the
  straight- and bent-stripes confgirations, respectively. Numbers
  associated to particles correspond to the particle positions in
  Table \ref{tab:straight} and Table \ref{tab:bent},
  respectively. (\textbf{e}) Parameters associated to every plaquette:
  $a$, $b$, $c$ and $\alpha$.}    
\label{fig:Shells}  
\end{figure}

A prototypical geometrically-confined system is composed of short-range
repulsive colloids confined in a slit pore of two plates. 
Varying density and plate separation, discontinuous phase transitions
between layered, buckled, rhombic and adaptive prism crystal
structures occur \cite{schmidt1996,schmidt1997,oguz2012}. 
In the case of a slit pore with a plates interdistance slightly larger
than a colloid diameter, when density approaches the close-packing
value $\rho_{cp}$, colloids, due to their free-volume-dominated free
energy, tend to touch opposite walls, giving rise to effective
antiferromagnetic interactions \cite{han2008,shokef2009}, and to
glassy dynamics \cite{zhou2017}.  
Multiple configurations corresponding to the same $\rho_{cp}$ can be
obtained by alternating straight stripes of up and down spheres
(Figure \ref{fig:Shells}a) or by any set of zigzagging stripes 
(Figure \ref{fig:Shells}b). 
This ground-state degeneracy implies a subextensive residual
entropy at $\rho_{cp}$ ($S_0\sim\sqrt{N}$, with N the number of
particles in the system) \cite{han2008} so that the residual entropy
per particle tends to zero in the thermodynamic limit.   
At $\rho=\rho_{cp}$, in the straight- or zigzagging-stripes
configuration, a colloidal sphere is surrounded by colloids 
touching the same wall (giving rise to a frustrated bond) and by
colloids touching the opposite wall (satisfied bond).
The terms frustrated and satisfied bonds refer both to the packing
consideration of neighboring spheres wanting to touch opposite walls,
and also from the obvious connection to an antiferromagnetic spin-1/2
Ising model \cite{shokef2009}. In this case, nearest
neighbors at opposite spin states satisfy the antiferromagnetic
interaction between them, while those at the same spin state represent
the frustration of this Ising model on the triangular lattice
\cite{wannier1950}.  
The three-dimensional (3D) network of links connecting the centers of
neighboring spheres is composed of tilted equilateral triangles,
since the 3D distance between the centers of each pair of contacting
spheres is equal to the sphere diameter. 
When the centers of the colloids are projected on the plane parallel
to the plates, it reduces to a 2D network composed of isosceles
triangles. In this network the longer link of each triangular
plaquette corresponds to a frustrated bond, while the two 
shorter links correspond to satisfied bonds. 
Straight and zigzagging stripes are the only configurations
corresponding to a 2D network of isosceles triangles which can tile
the plane \cite{han2008}. 

This geometric mechanism underlying the ground state of the
buckled colloidal system composed of straight or zigzagging stripes,
has been realised, experimentally also by packing a granular system in
a container under the effect of gravity \cite{harth2015,levay2018}, 
and theoreticelly with spins starting from the Wannier
antiferromagnetic Ising model on a triangular lattice
\cite{wannier1950} by allowing for the lattice to elastically deform
\cite{ilyin2009,shokef2011}.  
Zigzagging stripes patterns have been found also in the Ising model
on an anisotropic triangular lattice \cite{dublenych2013}. 
The Ising antiferromagnet on a deformable triangular lattice has the
same degeneracy of the ground state and subextensive entropy at $T=0$
as the colloidal system at $\rho=\rho_{cp}$ (that is
$S_0\sim\sqrt{N}$, equivalent to the $N^{2/3}$ scaling found in
Perovskite Oxynitrides \cite{camp2012}). In the following we refer to
$T=0$ as temperature being arbitrarily close to zero. 
Indeed, it has been shown that the third law of thermodynamics,
implying the unattainability of absolute zero temperature in a finite
number of steps and within a finite time, holds for arbitrary
classical or quantum systems or involving infinite-dimensional
reservoirs \cite{masanes2017}.

For temperature slightly larger than zero, the degeneracy is removed by
the order-by-disorder effect
\cite{villain1980,henley1987,henley1989,chubukov1992,reimers1993,bergman2007,patrykiejew2016,guruciaga2016}   
and in the elastic Ising model straight stripes represent the stable
phase \cite{shokef2011,leoni2017}, while bent stripes are selected in
the colloidal system for $\rho<\rho_{cp}$ when colloids are modeled
using hard or soft repulsive potentials \cite{leoni2017}. 
By tuning the attractive vs repulsive components of an asymmetric
power-law potential used to model colloids, we found that the stable
phase in the colloidal system can be turned from bent to straight
stripes for attraction large enough compared to repulsion 
\cite{leoni2017}.
We established a connection between the effect of the attraction on
the phase stability and the packing of hard spheres and their
entropy. We showed that other parameters of these systems are
irrelevant to the phase stability, as for example their
dimensionalities. Indeed, the elastic Ising antiferromagnet is defined
in 2D, while the colloidal system is 3D or quasi-2D due to the
buckling of the monolayer.  

In this paper we study a 2D isosceles triangular network which shows
the same ground-state degeneracy as the Ising elastic antiferromagnet
at $T=0$ or the colloidal monolayer at $\rho=\rho_{cp}$.
Increasing temperature above zero, the degeneracy is removed through
the order-by-disorder effect and we find the straight-stripes phase to
be selected for particles linked with harmonic interactions, while the
bent-stripes phase is more stable if only the repulsive component of
the harmonic inter-particle potential is considerd (that we call
repulsive harmonic potential, described in detail in the following 
section).    
This result suggests that the inversion of the stable phase by adding
attraction to repulsivly-interacting particles in triangular networks
is a general mechanism, irrespective of the dimensionality of
the system.

\section{Isosceles triangular network model}

The model we consider is composed of particles in a 2D triangular
network linked with springs of two different lengths at rest such that
every plaquette or triangle is formed by a longer edge $2a$ and two
shorter identical edges $c$, and thus has a height $b=\sqrt{c^2-a^2}$
and a head angle $\alpha$ such that $a=b\tan(\alpha/2)$ (see
Figure \ref{fig:Shells}e).  
We will first consider particles interacting through a harmonic
potential and then will consider a repulsive harmonic potential, as
defined below. The Hamiltonian of the system for harmonic
inter-particle interaction can be written as  
\begin{equation}
\mathcal{H}^h=\sum_{m,n}\sum_{l=1}^3\dfrac{K}{2}\left(dr_l-dr_0\right)^2,
\end{equation}
where $K$ is the spring constant, which is assumed to be identical for
all springs, $1\leq m,n\leq L$, $N=L^2$ is the
number of particles or nodes of the network and the index $l$ runs
over three of the six neighbors each particle has to avoid double
counting of bonds. The positions of the particles are described by the
coordinates $\{x_i,y_i\}$.
At $T=0$ multiple degenerate states, represented by straight or
zigzagging stripes, minimize the system free energy.  
For $T>0$ the system cannot jump from one configuration to another,
but at $T=0$ it is at mechanical equilibrium in every state under
consideration. Therefore, the present model is not ergodic by
construction. In the conclusion we will discuss entropy calculation in
non ergodic systems. 
In order to study the stability of the system at $T>0$, we will
consider small fluctuations about the equilibrium position 
described by small displacements $\{u_i,v_i\}$ of all particles in
the straight and bent configurations.
$dr$ is the distance between particles $i$ and $j$, and its square is
thus given by  
\begin{equation}\label{equ:dr}
dr^2=(dx+du)^2+(dy+dv)^2=dr_0^2+2(dxdu+dydv)+du^2+dv^2
\end{equation}
where $dx=x_i-x_j$, $dy=y_i-y_j$, $du=u_i-u_j$, $dv=v_i-v_j$ and
$dr_0=(dx^2+dy^2)^{1/2}$ is the length at rest of the spring linking
particles $i$ and $j$, which can take the values $c$ or $2a$ (see
Figure \ref{fig:Shells}e and Tables \ref{tab:straight},
\ref{tab:bent}).  
Since we consider the expansion around mechanical equilibrium, we
will drop terms linear in $du$ and $dv$ and write:
$dr^2=dr_0^2+du^2+dv^2$.  
Expanding to harmonic order the expression of $dr$ that we get from
Eq.(\ref{equ:dr}), we obtain  
\begin{equation}
dr=dr_0+\dfrac{du^2}{2dr_0}\left(1-\dfrac{dx^2}{dr_0^2}\right)+\dfrac{dv^2}{2dr_0}\left(1-\dfrac{dy^2}{dr_0^2}\right)-\dfrac{dxdydudv}{dr_0^3}.
\end{equation}
The Hamiltonian of the straight-stripes configuration, with particle
positions specified for the unit-cell in Table \ref{tab:straight}, is
\begin{equation}
\mathcal{H}_s^h=\dfrac{K}{2}\sum_{m,n}\left[du_1^2+\dfrac{a^2}{c^2}(du_2^2+du_3^2)+\dfrac{b^2}{c^2}(dv_2^2+dv_3^2)-\dfrac{2ab}{c^2}(-du_2dv_2+du_3dv_3)\right].
\end{equation}
Using the relations $du_l=u_l-u_0$ and $dv_l=v_l-v_0$ we get
\begin{equation}
\begin{array}{ll}
\mathcal{H}_s^h= &
K\sum_{m,n}\left[u_0^2-u_0u_1+\dfrac{a^2}{c^2}(2u_0^2-u_0u_2-u_0u_3)+\dfrac{b^2}{c^2}(2v_0^2-v_0v_2-v_0v_3)\right.\\
&\\
&\left.-\dfrac{2ab}{c^2}(u_0v_2+u_2v_0-u_0v_3-u_3v_0)\right].
\end{array}
\end{equation}
The Hamiltonian of the bent-stripes configuration, with particle
positions specified for the unit-cell in Table \ref{tab:bent}, is
\begin{equation}
\begin{array}{ll}
\mathcal{H}_b^h= & \dfrac{K}{2}\sum_{t,n}\left\{du_{10}^2+\cos^2\alpha
  du_{50}^2+\sin^2\alpha dv_{50}^2-\sin(2\alpha)
  du_{50}dv_{50}+\sin^2(\dfrac{\alpha}{2})
  (du_{40}^2+du_{41}^2+du_{31}^2)\right.\\
&\\
& +\sin^2(\dfrac{3\alpha}{2})du_{21}^2+\cos^2(\dfrac{\alpha}{2})
  (dv_{40}^2+dv_{41}^2+dv_{31}^2)+\cos^2(\dfrac{3\alpha}{2})
  dv_{21}^2\\
&\\
&\left.-\sin\alpha(-du_{40}dv_{40}+du_{41}dv_{41}-du_{31}dv_{31})+\sin(3\alpha) 
  du_{21}dv_{21}\right\},
\end{array}
\end{equation}
where $1\leq t\leq L/2$. Indeed, the unit-cell of the bent stripe
configuration includes two particles (see Figure \ref{fig:Shells}d):
particle 0, which represents the particles with odd $m$, and particle
1, which represents the particles with even m. Therefore, we set for
particle 0, $m=2t-1$, and for particle 1, $m=2t$.
Using the relations $du_{l0}=u_l-u_0$, $du_{l1}=u_l-u_1$ and
$dv_{l0}=v_l-v_0$, $dv_{l1}=v_l-v_1$ we get
\begin{equation}
\begin{array}{ll}
\mathcal{H}_b^h= &
\dfrac{K}{2}\sum_{t,n}\left\{\left[2-\sin^2\alpha+3\sin^2(\dfrac{\alpha}{2})+\sin^2(\dfrac{3\alpha}{2})\right](u_0^2+u_1^2)+(\sin\alpha-\sin2\alpha+\sin3\alpha)(u_0v_5\right.\\ 
&\\
& +u_5v_0)-\sin\alpha(u_0v_4+u_4v_0-u_1v_4-u_4v_1+u_1v_3+u_3v_1)-\sin3\alpha(u_1v_2+u_2v_1)\\
&\\
& +\left[1-\cos^2\alpha+3\cos^2(\dfrac{\alpha}{2})+\cos^2(\dfrac{3\alpha}{2})\right](v_0^2+v_1^2)-2\sin^2\alpha  
  v_0v_5-2\cos^2(\dfrac{\alpha}{2})(v_0v_4+v_1v_3\\
&\\
& \left. +v_1v_4)-2\cos^2(\dfrac{3\alpha}{2})v_1v_2\right\}. 
\end{array}
\end{equation}

\begin{figure}[h!]
\centering
\includegraphics[clip=true,width=13cm,height=9.1cm]{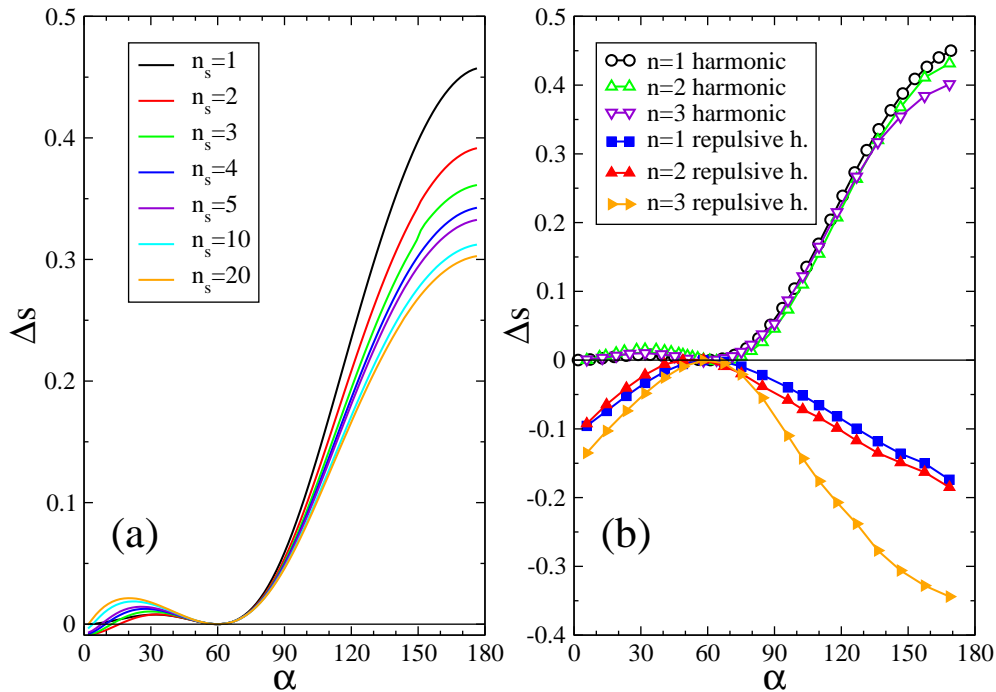}
\caption{(\textbf{a}) Entropy difference per particle $\Delta s$,
  between straight- and bent-stripe configurations vs deformation
  angle $\alpha$ for increasing numbers of shells $n_s$ of fluctuating
  particles.  
(\textbf{b}) $\Delta s$ vs $\alpha$ for $n=1,2,3$ particles free to
  move for harmonic (open symbols) and repulsive harmonic (filled
  symbols) interactions.}  
\label{fig:Ds}  
\end{figure}

For a harmonic interparticle potential, the Hamiltonian for straight
and bent stripes configurations we expanded around mechanical
equilibrium takes the quadratic form: 
$\mathcal{H}=K\sum_{m,n}A_{m,n}q_mq_n$, where $\{q\}=\{u,v\}$
represents small displacements about the equilibrium position of every
particle.   
In the canonical ensemble the difference between the entropy per
particle of the straight and bent configurations for such 
Hamiltonian is \cite{leoni2017}: $\Delta 
s=(S_s-S_b)/N=1/(2N)\ln(\|A_b\|/\|A_s\|)$ where the subscript $s$ refers
to straight and $b$ to bent, and $\|A\|$ is the determinant of A. 
The dimensionless matrix A depends only on the deformation angle
$\alpha$ and on the zigzagging-stripe realization.
In Ref.\cite{leoni2017} we used a recursive method to obtain the
matrix A in the case of the elastic Ising model for any subset of the
network composed by shells of particles (see Figure
\ref{fig:Shells}). Here we apply the same method to the 2D spring 
network model. The number of particles $n$ belonging to the shells up
to $n_s$ is given by $n=1+3n_s(n_s-1)$. In our shell-expansion
calculation, these $n$ particles are free to move, while the other
$N-n$ particles of the network are frozen in their equilibrium
position. Increasing $n$, $\Delta s$ should converge to the exact
result (see Figure \ref{fig:Ds}a), which includes the simultaneous
fluctuation of all particles in the system.  
In Figure \ref{fig:Ds}a we show $\Delta s$ for the 2D harmonic network
model for a number of shells up to $n_s=20$, that is $n=1141$
particles free to move. From it we can see that $\Delta s>0$ for every
deformation angle $\alpha$ of the network and for every order $n_s$ of
the expansion (except for very small deviations at small $\alpha$ and
small $n_s$). Considering only one particle free to 
move, $n_s=1$, gives a qualitative indication on the behavior of
$\Delta s$ for every orders of approximation.
This rapid convergence with increasing $n_s$ gives confidence in this
expansion method when we will apply it for purely repulsive
interactions, for which we are technically much more limited in the
number of particles that we may numerically calculate the simultaneous
fluctuation of. 

Now we consider the same 2D triangular network model, but for
repulsive harmonic interactions, that is we consider the following
Hamiltonian 
\begin{equation}
\mathcal{H}^r=\sum_{m,n}\sum_{l=1}^3\dfrac{K}{2}\left(dr_l-dr_0\right)^2\theta(dr_0-dr_l),
\end{equation}
where $\theta()$ is the Heaviside step function. Namely, now each
spring applies a restoring force only when compressed ($dr_l<dr_0$)
and there is no resistance to stretching ($dr_l>dr_0$).  
In this case we have to numerically integrate the partition function
in order to get the entropy of straight- and bent-stripe
configurations. 
In Figure \ref{fig:Ds}b we show $\Delta s$ for repulsive
harmonic interactions for $n=1,2,3$. Numerical calculations for $n>3$
are beyond our computational reach (numerical integration for a
number of variables larger than 6, in our case suffers from
fluctuating results for a limited capacity in the precision).   
For $n=1$ particle free to move,
each particle can be equivalently chosen to be free. For $n=2$
we consider the particles 0 and 1 (see
Figure \ref{fig:Shells}c,d). For $n=3$ we consider the particles 0,
1, 2 and 0, 1, 4 for the straight and bent configurations,
respectively (see Figure \ref{fig:Shells}c,d).  
As for the 3D spring network of Ref.\cite{leoni2017}, we find that
also in our 2D model $\Delta s<0$ for repulsive interactions.
\begin{table}
\caption{Distances between the neighboring particles and the central
  particle 0 in the unit cell of the straight-stripe
  configuration. Particle positions are graphically shown in Figure 
  \ref{fig:Shells}c.} 
\centering
\begin{tabular}{|c|*{3}{c|}}
\hline
particles & dx & dy & $dr_0$\\
\hline
1,0 & $2a$ & 0 & $2a$\\

2,0 & $a$  & $b$ & $c$\\

3,0 & $-a$ & $b$ & $c$\\ 

4,0 & $-2a$  & 0 & $2a$\\

5,0 & $-a$ & $-b$ & $c$\\

6,0 & $a$ & $-b$ & $c$\\
\hline
\end{tabular}
\label{tab:straight}
\end{table}
\begin{table}
\caption{Distances between the neighboring particles and $0$ and $1$ 
particles in the unit cell of the maximally zigzagging-stripe
configuration. Particle positions are graphically shown in Figure 
  \ref{fig:Shells}d.}  
\centering
\begin{tabular}{|c|*{3}{c|}}
\hline
particles & dx & dy & $dr_0$\\
\hline
1,0  & $2a$ & 0 & $2a$\\

4,0  & $a$ & $b$ & c\\

5,0  & $2a\left(1-8\dfrac{a^2}{c^2}\right)$ &
$4b\left(1-\dfrac{b^2}{c^2}\right)$ & $2a$\\

6,0  & $a\left(1-4\dfrac{b^2}{c^2}\right)$ &
$b\left(3-4\dfrac{b^2}{c^2}\right)$ & $c$\\

7,0  & $-a$ & $-b$ & $c$\\

8,0  & $a$  & $-b$ & $c$\\

2,1  & $-a\left(1-4\dfrac{b^2}{c^2}\right)$ &
$-b\left(3-4\dfrac{b^2}{c^2}\right)$ & $c$\\

3,1  & $a$ & $b$ & $c$\\

4,1  & $-a$ & $b$ & $c$\\

8,1  & $-a$  & $-b$ & $c$\\

9,1  & $-2a\left(1-2\dfrac{b^2}{c^2}\right)$ &
$-4b\left(1-\dfrac{b^2}{c^2}\right)$ & $2a$\\
\hline
\end{tabular}
\label{tab:bent}
\end{table}
For the case $n=1$ we can show how the inversion of the stable
phase when turning from harmonic to repulsive harmonic potential
depends on the spatial configurations of straight and bent stripes,
and in particular of the angular distribution of neighboring particles
around each particle in the network. 

\begin{figure}[h!]
\centering
\includegraphics[clip=true,width=8cm]{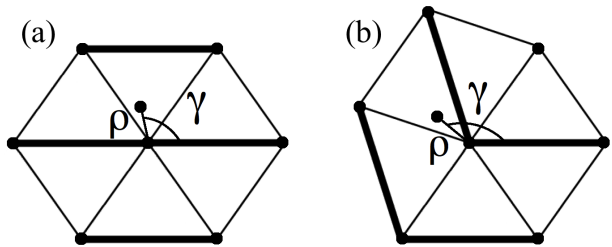}
\caption{Example of straight (\textbf{a}) and bent (\textbf{b})
  configuration for $n=1$ particle free to move. The deviation of the
  central particle from its equilibrium position is described by polar
  coordinates ($\rho,\gamma$).}    
\label{fig:Stripes0}  
\end{figure}

For $n=1$ the computation of the canonical partition function for the
repulsive harmonic system can be easily reduced to the integration of
single variable functions using polar coordinates $(\rho,\gamma)$ (see
Figure \ref{fig:Stripes0}). 
The Hamiltonian of the free particle 0 can be written as
$\mathcal{H}_0^r=1/2K\rho^2\sum_{i=1}^6g_i(\alpha,\gamma)$ where
$\rho^2g_i(\alpha,\gamma)$ is the contribution to $\mathcal{H}_0^r$
coming from the neighbor $i$ of the particle 0, and the function
$g_i$ depends on the coordinates of the particle $i$, as specified
below. The canonical partition function is thus
\begin{equation}\label{equ:Z0}
Z_0^r=\int_{0}^{\infty}\int_{0}^{2\pi}\exp\left[-\beta\dfrac{K}{2}\rho^2\sum_{i=1}^6g_i(\alpha,\gamma)\right]\rho
d\rho d\gamma=\dfrac{1}{\beta
  K}\int_{0}^{2\pi}\dfrac{d\gamma}{\sum_{i=1}^6g_i(\alpha,\gamma)}=\dfrac{I(\alpha)}{\beta
K}
\end{equation}
where $\beta=1/(K_BT)$ is the Boltzmann factor and
$I(\alpha)=\int_{0}^{2\pi}[\sum_{i=1}^{6}g_i(\alpha,\gamma)]^{-1}d\gamma=\int_{0}^{2\pi}f^{-1}(\alpha,\gamma)d\gamma$
with $f(\alpha,\gamma)=\sum_{i=1}^{6}g_i(\alpha,\gamma)$.  
In this case we have $\Delta s=\ln(I_s(\alpha)/I_b(\alpha))$.
The difference between the calculation of the harmonic and the
repulsive harmonic partition function is that in the former case the
functions $g_i$ contributes to the integral $I(\alpha)$ for any angle
$0\leq\gamma\leq 2\pi$, while in the latter case every $g_i$
contributes to $I(\alpha)$ for a specific range of angles only.
For the repulsive harmonic potential in the straight-stripes
configuration we need to consider the azimuthal ranges coming from
each one of the neighboring particles: 
\begin{equation}
\begin{array}{ll}
f_s^r(\alpha,\gamma)= &
\cos^2\gamma\Big[\theta(\gamma)\theta(\pi/2-\gamma)+\theta(\gamma-3\pi/2)\theta(2\pi-\gamma)\Big]+\sin^2(\alpha/2+\gamma)\Big[\theta(\gamma)\theta(\pi-\alpha/2-\gamma)\\
&\\
&
  +\theta(\gamma-2\pi+\alpha/2)\theta(2\pi-\gamma)\Big]+\sin^2(\alpha/2-\gamma)\Big[\theta(\gamma-\alpha/2)\theta(\pi+\alpha/2-\gamma)\Big]\\
&\\
&
+\cos^2\gamma\Big[\theta(\gamma-\pi/2)\theta(3\pi/2-\gamma)\Big]+\sin^2(\alpha/2+\gamma)\Big[\theta(\gamma-\pi+\alpha/2)\theta(2\pi-\alpha/2-\gamma)\Big]\\
&\\
&+\sin^2(\alpha/2-\gamma)\Big[\theta(\gamma-\pi-\alpha/2)\theta(2\pi-\gamma)+\theta(\gamma)\theta(\alpha/2-\gamma)\Big].
\end{array}
\end{equation}
Due to the symmetry of the straight-stripes configuration, thanks to
which the reflection about the origin of each neighbor transforms it
in another neighboring particle (see Figure \ref{fig:Shells}c and
Table \ref{tab:straight}), we can write the function $f_s^r$ by just
taking the contribution of every $g_i$ without the condition imposed
by the Heaviside step functions and dividing it by 2, that is 
\begin{equation}   
f_s^r(\alpha,\gamma)=\cos^2\gamma+\sin^2(\alpha/2+\gamma)+\sin^2(\alpha/2-\gamma)=\cos^2\gamma(2-\cos\alpha)+\cos^2(\alpha/2)
\end{equation}

For the repulsive harmonic potential in the bent configuration we need
to consider the azimuthal ranges coming from each one of the
neighboring particles: 
\begin{equation}
\begin{array}{ll}
f_b^r(\alpha,\gamma)= &
\cos^2\gamma\Big[\theta(\gamma)\theta(\pi/2-\gamma)+\theta(\gamma-3\pi/2)\theta(2\pi-\gamma)\Big]+\sin^2(\alpha/2+\gamma)\Big[\theta(\gamma)\theta(\pi-\alpha/2-\gamma)\\
&\\
&
  +\theta(\gamma-2\pi+\alpha/2)\theta(2\pi-\gamma)\Big]+\cos^2(\alpha+\gamma)\Big[\theta(\gamma-\pi/2+\alpha)\theta(3\pi/2-\alpha-\gamma)\Big]\\
&\\
&
+\sin^2(3\alpha/2+\gamma)\Big[\theta(\gamma-\pi+3\alpha/2)\theta(2\pi-3\alpha/2-\gamma)\Big]+\sin^2(\alpha/2+\gamma)\Big[\theta(\gamma-\pi+\alpha/2)\\
&\\
&\cdot\theta(2\pi-\alpha/2-\gamma)\Big]+\sin^2(\alpha/2-\gamma)\Big[\theta(\gamma-\pi-\alpha/2)\theta(2\pi-\gamma)+\theta(\gamma)\theta(\alpha/2-\gamma)\Big]
\end{array}
\end{equation}
%
\begin{figure}[h!]
\begin{center}
\begin{minipage}{5.2cm}
\includegraphics[clip=true,width=5.2cm,height=3.9cm]{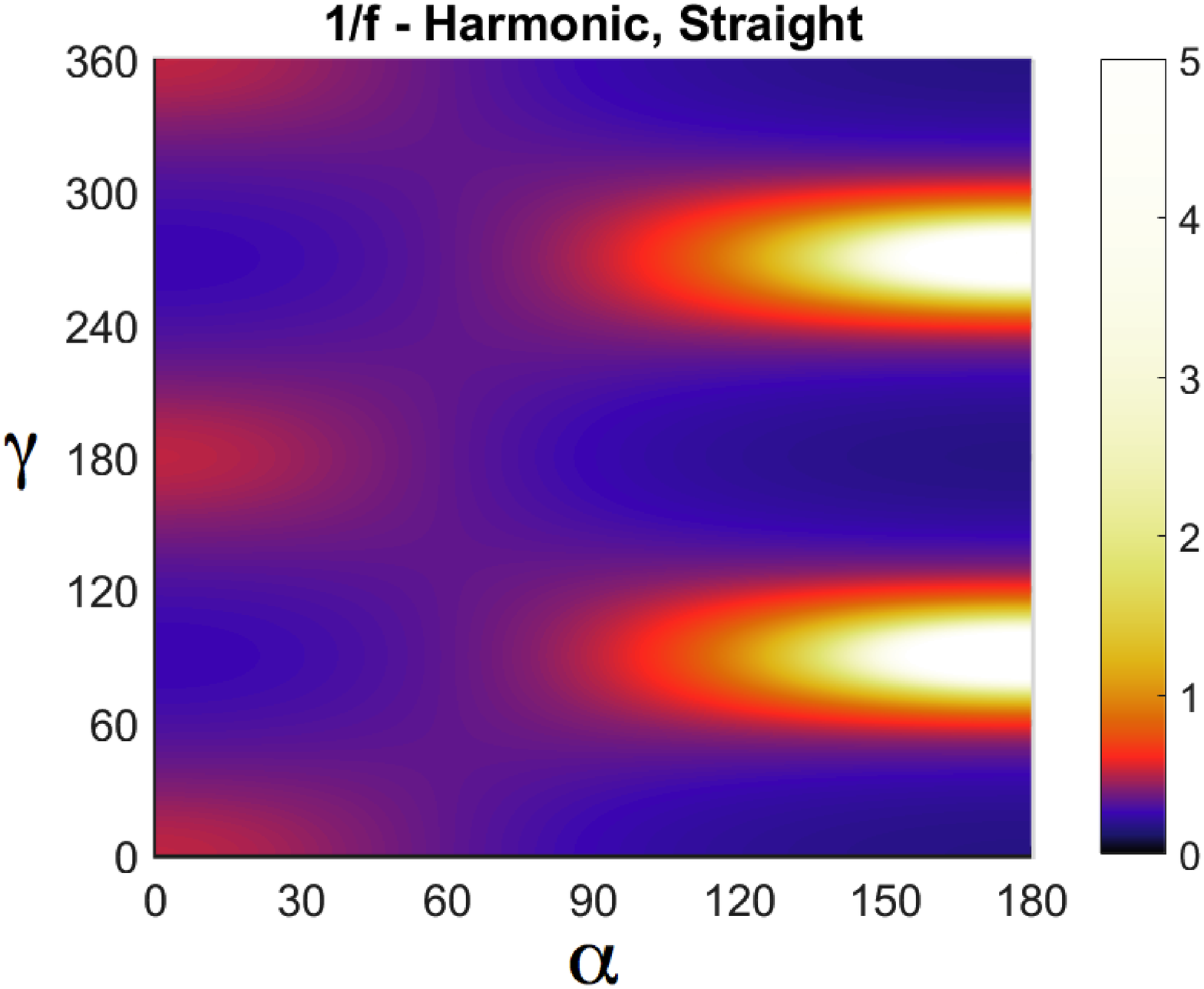}
\includegraphics[clip=true,width=5.2cm,height=3.9cm]{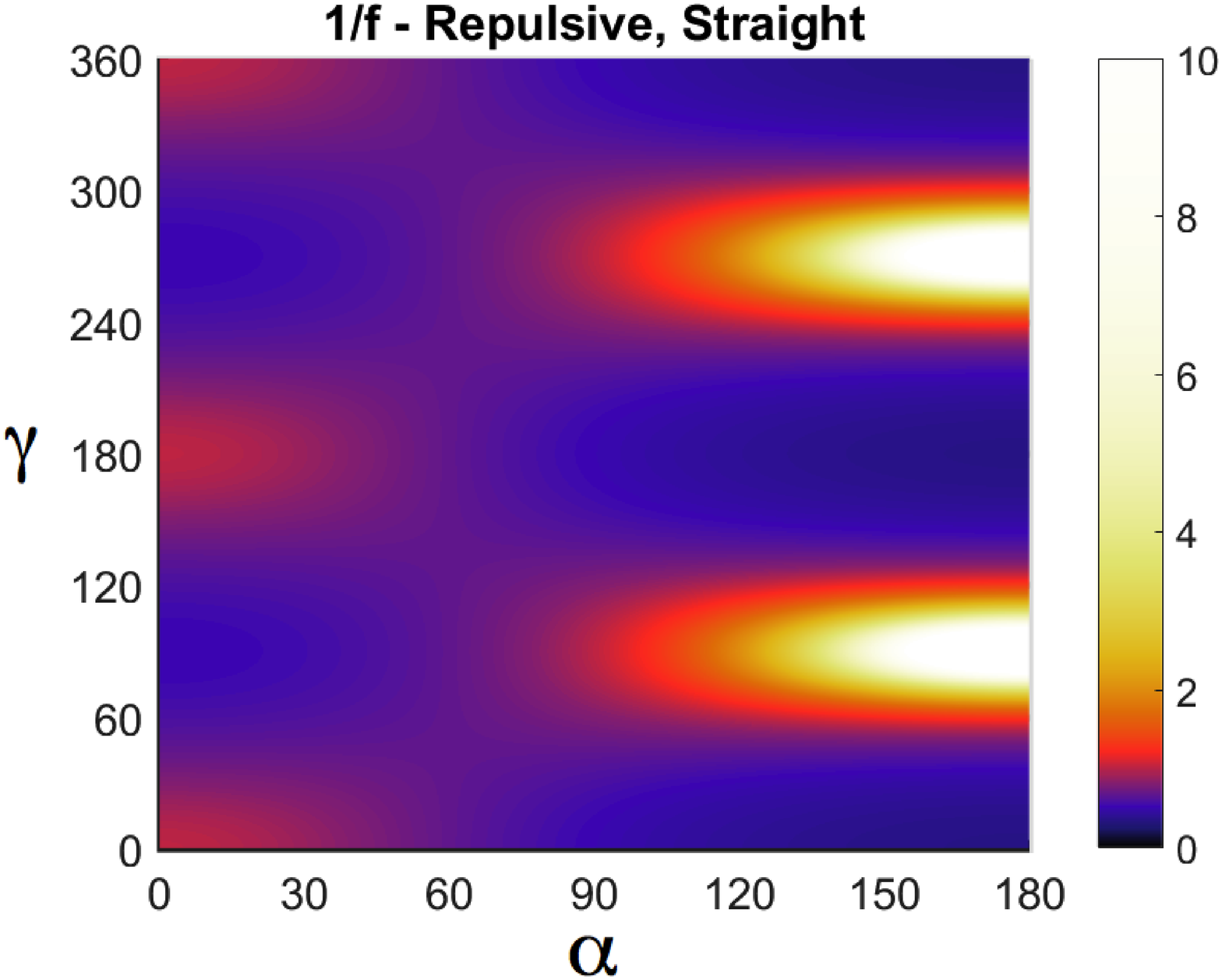}
\end{minipage}
\begin{minipage}{5.2cm}
\includegraphics[clip=true,width=5.2cm,height=3.9cm]{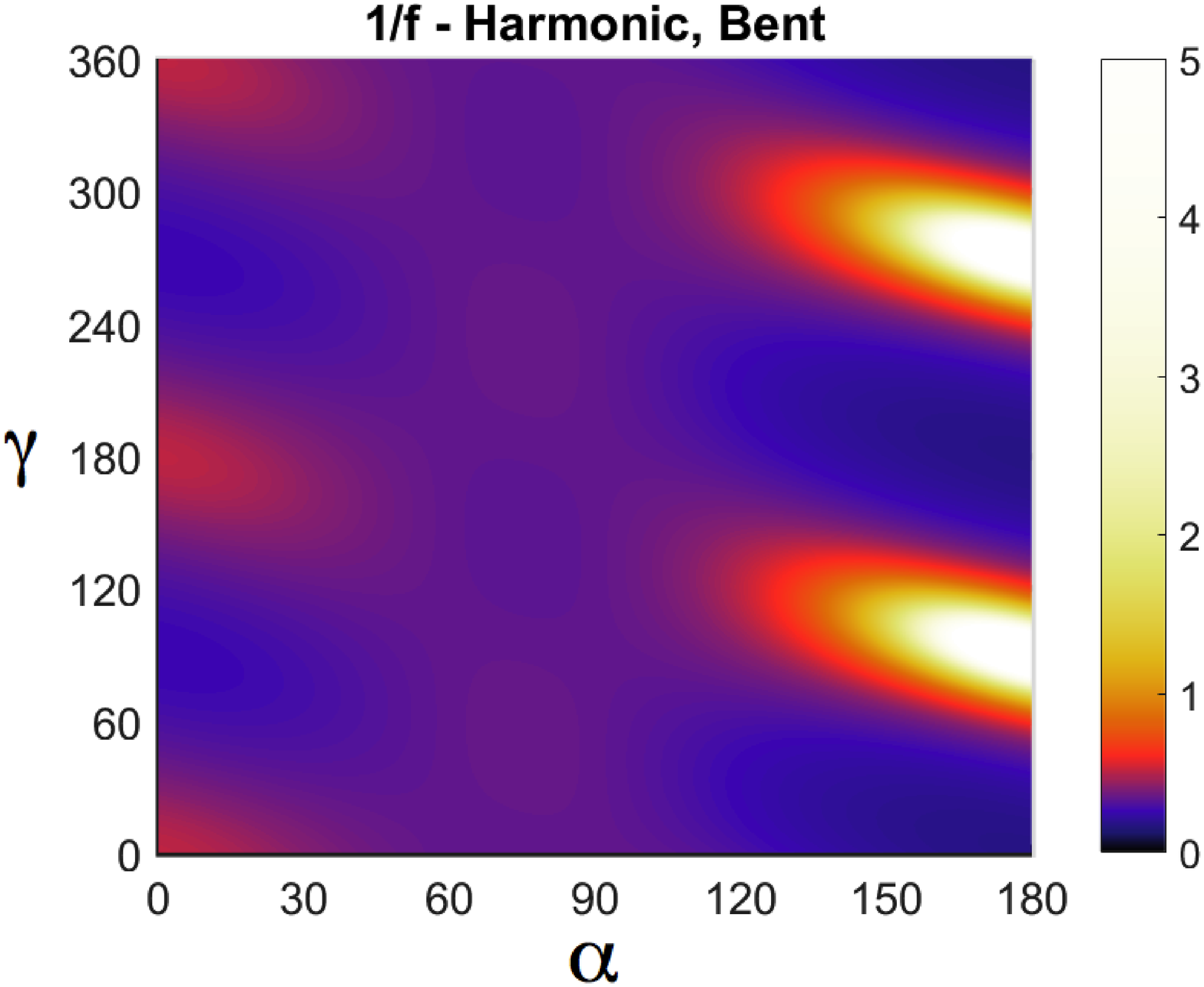}
\includegraphics[clip=true,width=5.2cm,height=3.9cm]{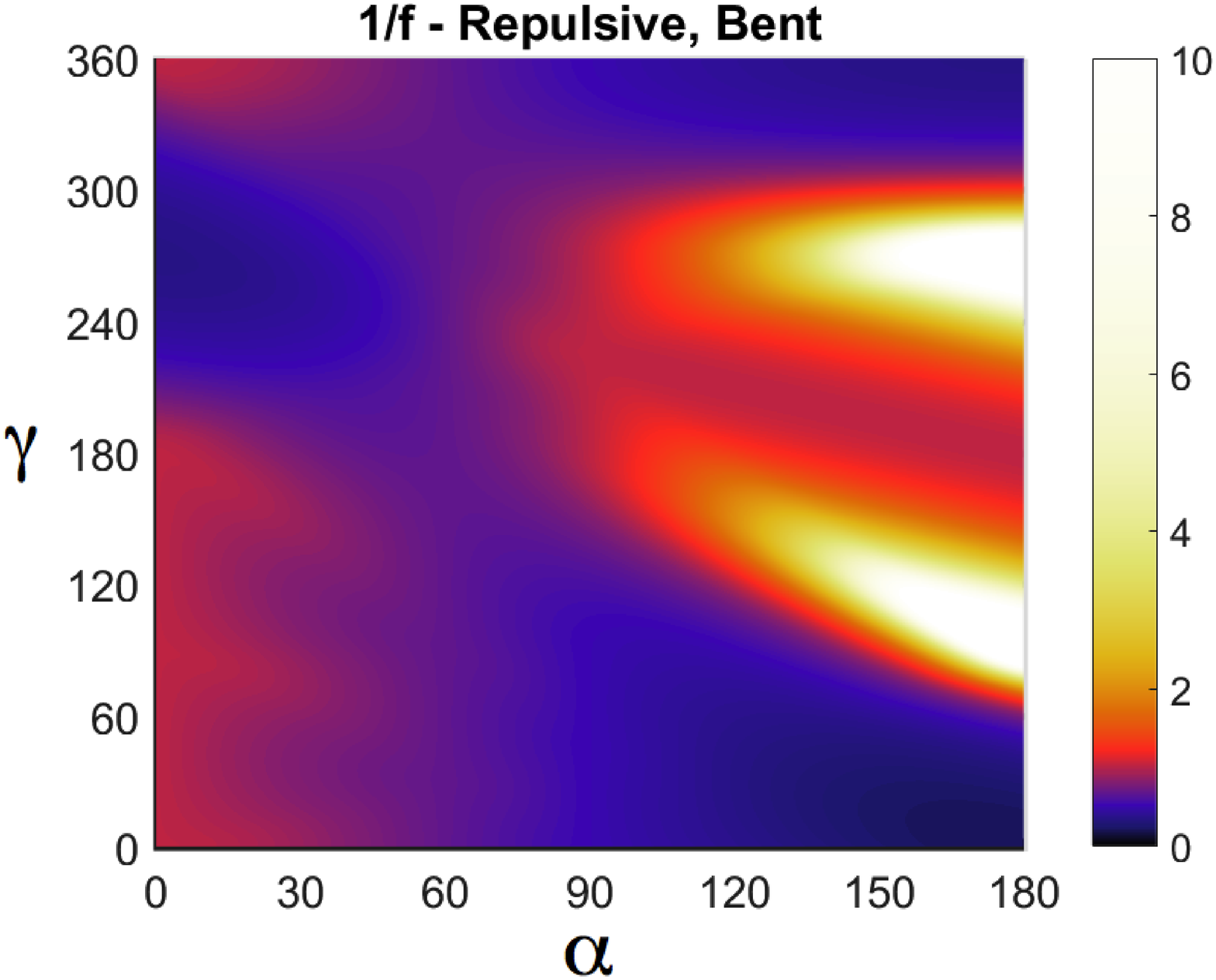}
\end{minipage}
\caption{Colour diagram of $1/f$, as defined in the text, as a
  function of the angles $\alpha$ and $\gamma$, for harmonic and
  repulsive harmonic interactions and for straight- and bent-stripes 
  configurations.}  
\label{fig:1overf}  
\end{center}
\end{figure}
In Figure \ref{fig:1overf} we show $1/f$ for hamonic and repulsive
harmonic interactions for straight and bent configurations as a
funciont of the lattice deformation angle $\alpha$ and the azimuthal
direction in space $\gamma$, over which we numerically integrate in
order to get the value of the function $I$ for that specific angle
$\alpha$, which in turn sets the entropy via Equation \ref{equ:Z0}.  
From Figure \ref{fig:1overf} we can see, particularly for big angles
$\alpha$ for which $\Delta s$ takes its larger values (see Figure
\ref{fig:Ds}b), that repulsion accentuates the contribution to the 
function $I$ for bent stripes (corresponding in Figure \ref{fig:1overf}
to a wider region composed of brighter colors, i.e. white, yellow and
red, for repulsive harmonic over harmonic interaction in the case of
bent stripes). More in general, we can say that repulsion accentuates
differences in the contribution to the partition function and thus to
the free energy between symmetric and asymmetric distribution of
neighboring particles.

\section{Conclusions}

We studied a 2D triangular-network model composed of particles
interacting through harmonic or repulsive harmonic springs.
At $T=0$ the ground state is degenerate and composed of straight or
any set of zigzagging-stripes configurations. 
At $T>0$ we found that the stable phase is composed of straight or
bent stripes depending on the harmonic or repulsive harmonic nature of 
particle interaction, respectively.
This selection mechanism of the stable phase through the
order-by-disorder effect is equivalent to that observed in the
colloidal \cite{leoni2017} and Ising \cite{shokef2011}
antiferromagnets irrespective of the dimensionality of the system.
This suggests that the phase inversion of isosceles triangular
networks is controlled by the attraction component of the
interparticle interaction. We suggest that this is due to the fact
that repulsive interactions accentuate differences in the contribution
to the free energy between symmetric and asymmetric distribution of
neighboring particles, as we have shown for $n=1$ free particle
calculation. 

Both the Ising antiferromagnet on a deformable triangular lattice and
the 2D isosceles triangular network model at $T=0$ from one side, and
the colloidal monolayer at $\rho=\rho_{cp}$ from the other side, have
the same degeneracy and subextensive entropy $S\sim\sqrt{N}$ and thus
a vanishing residual entropy per particle in the thermodynamic limit. 
At $T>0$ for the triangular networks and at $\rho<\rho_{cp}$ for the 
colloidal monolayer this degeneracy is removed, but they can still
have a residual entropy per particle for finite system size if the
ergodicity is broken.
Indeed, even at $T>0$ or $\rho<\rho_{cp}$ a system may be trapped in a
local minimum of the free-energy landscape and thermal fluctuations
are not large enough for a small system to overcome energy barriers.
For ergodic systems the time average of observables can be computed by
using ensemble averages thanks to the Birkhoff theorem
\cite{birkhoff1931}. From the other hand, for non-ergodic systems, the
phase space is divided into disjoined sets. In this case, states can
be counted either following the kinetic view \cite{mauro2007}, for
which only states visited by the system at the observational time
scale are taken into account, or following the Edwards approach
\cite{edwards1989}, for which all possible states are considered
regardless of whether they are explored or not by the
system. Recently, the Edwards hypothesis has been proved to be valid
at the un-jamming point \cite{martiniani2017}. In thermal ergodic
systems at equilibrium, the two sampling methods give the same result. 
Following the Edwards approach we can conclude that an indication of
the presence of residual entropy in a system is given by the
ergodicity breaking (which can be checked for generic temperature or
density) instead of by the degeneracy of the ground state (defined for
$T=0$ or $\rho=\rho_{cp}$ only). 

\acknowledgments{This research was supported by the Israel Science
  Foundation Grant No. 968/16.} 

\bibliography{Entropy_Leoni-Shokef.bib}

\end{document}